\documentclass[pra,aps,twocolumn,superscriptaddress,showpacs]{revtex4}

\usepackage[tbtags]{amsmath}
\usepackage{amssymb}
\usepackage{amsfonts}
\usepackage[dvips]{graphicx,color}
\usepackage{amsmath}
\usepackage{amssymb}
\usepackage[dvips]{graphicx}
\usepackage{graphicx}
\usepackage{dcolumn}
\usepackage{bm}
\usepackage{setspace}
\usepackage{amsfonts,color}
\usepackage{rotating}
\usepackage{makeidx}
\usepackage{amsthm}
\usepackage{xcolor}

\newcommand{\tr}{\mbox{Tr}}

\begin{document}
\title{Beating the Clauser-Horne-Shimony-Holt and the Svetlichny games with Optimal States}

\author{Hong-Yi~Su}
 \affiliation{Theoretical Physics Division, Chern Institute of Mathematics, Nankai University,
 Tianjin 300071, People's Republic of China}
  \affiliation{Department of Physics Education, Chonnam National University, Gwangju 500-757, Republic of Korea}

\author{Changliang Ren}
\email{renchangliang@cigit.ac.cn}
 \affiliation{Center of Quantum information, Chongqing Institute of Green and Intelligent Technology, Chinese Academy of Sciences, People's Republic of China}

\author{Jing-Ling~Chen}
 \email{chenjl@nankai.edu.cn}
 \affiliation{Theoretical Physics Division, Chern Institute of Mathematics, Nankai University,
 Tianjin 300071, People's Republic of China}
 \affiliation{Centre for Quantum Technologies, National University of Singapore,
 3 Science Drive 2, Singapore 117543}

\author{Fu-Lin Zhang}
\email{flzhang@tju.edu.cn}
\affiliation{Physics Department, School of Science, Tianjin
University, Tianjin 300072, China}

\author{Chunfeng Wu}
\affiliation{Pillar of Engineering Product Development, Singapore University of Technology and Design, 8 Somapah Road, Singapore 487372}

\author{Zhen-Peng Xu}
 \affiliation{Theoretical Physics Division, Chern Institute of Mathematics, Nankai University,
 Tianjin 300071, People's Republic of China}

\author{Mile Gu}
\affiliation{Centre for Quantum Technologies, National University of
Singapore, 3 Science Drive 2, Singapore 117543}

\author{Sai Vinjanampathy}
\affiliation{Centre for Quantum Technologies, National University of
Singapore, 3 Science Drive 2, Singapore 117543}

\author{L. C. Kwek}
 \affiliation{Centre for Quantum Technologies, National University of Singapore,
 3 Science Drive 2, Singapore 117543}
 \affiliation{National Institute of Education and Institute of Advanced Studies,
 Nanyang Technological University, 1 Nanyang Walk, Singapore 637616}

\begin{abstract}

We study the relation between the maximal violation of Svetlichny's inequality and the mixedness of quantum states and obtain the optimal state (i.e., maximally nonlocal mixed states, or MNMS, for each value of linear entropy) to beat the Clauser-Horne-Shimony-Holt and the Svetlichny games.   For  the two-qubit and three-qubit MNMS, we showed that these states are also the most tolerant state against white noise, and thus serve as valuable quantum resources for such games. In particular,
the quantum prediction of the MNMS decreases as the linear entropy increases, and then ceases to be nonlocal when the linear entropy reaches the critical points ${2}/{3}$ and  ${9}/{14}$ for the two- and three-qubit cases, respectively. The MNMS are related to classical errors in experimental preparation of maximally entangled states.

\end{abstract}

\pacs{
03.65.Ud, 03.67.Mn, 02.50.Le}

\maketitle
\section{Introduction}

Arising initially from the debate on the incompleteness of quantum
mechanics~\cite{EPR}, quantum nonlocality or more correctly speaking, a nonlocal realistic description of Nature, is now a valuable resource in many aspects
of quantum information science~\cite{Acin07,Pironio10}. Quantum
nonlocality is witnessed by the violation of Bell-type inequalities, and these inequalities generally admit local-hidden-variable (LHV)
models~\cite{Bell64,chsh69}, and they arguably provide some of the most intriguing features of quantum mechanics.

There have been many investigations on Bell-type inequalities for quantum systems
of arbitrary parties and dimensions~\cite{MABK1,MABK2,MABK3,WWZB1,WWZB2}. Inequalities involving many-body correlations are important since such correlations dominate the condensed matter of many-body physics. For multipartite
systems, the issue of quantum nonlocality is rather subtle.  One such subtlety arises naturally within the context of determining if the nonlocality
of an $N$-qubit system is intrinsically related to genuine $N$-qubit correlations, or just simply a convex combinations of nonlocal
correlations within subsystems. Another interesting question pertains to whether  quantum theory always admits nonlocal features and whether certain subsystems have well-defined properties.  Therefore, it is important to carry out tests for many-body scenario and see if it is ``immune to any explanation in terms of mechanisms involving fewer bodies"~\cite{Svetlichny}.   Genuine multiparitite entanglement is first explored by Svetlichny~\cite{Svetlichny}
in 1987 where he constructed  a family of Bell-type inequalities, renowned now as Svetlichny's inequalities,  for a three-qubit system  from  hybrid local-nonlocal hidden variable models. Violation of such inequalities immediately leads to genuine multipartite nonlocality. In Svetlichny's inequalities, all subsystems  necessarily participate, with no single subsystem possessing distinctive well-defined properties.

Svetlichny's inequality  (SI) is now widely regarded as a useful tool for detecting genuine three-qubit nonlocality.  Aside from multipartite scenarios, it has also been extensively studied
for arbitrarily-dimensional systems~\cite{Svetlichny2,Svetlichny3,Gisin11,us11,acin2012,lee2013,adesso2014}. Note that
genuine multipartite nonlocality is not the same as genuine
multipartite entanglement (i.e., full entanglement).  The
latter describes the mathematical impossibility of separating a
quantum state into two parts. Put simply, nonlocality and entanglement serve as differential resources and they are both useful for different applications in quantum information science.

Unlike the typical Bell-type inequalities, there has been less research done on SI. There has been little understanding on SI with mixed states.
Exploring systems with mixed state is essential since environment-induced noise is in general unavoidable in real
experiments.  However, with mixed states,  it is generally harder to optimize the use of a quantum resource with respect to a
given measure of mixedness or purity.  There are however some interesting examples like  \emph{maximally entangled mixed states} (MEMS) and \emph{maximally discordant mixed states} (MDMS)
\cite{MEMS,MEMS1,MDMS}. Moreover, the borders between nonlocality, entanglement, and mixedness of states are not fully characterized yet. Studies on their differences and their inter-relations may reveal  insights for a better understanding of quantum theory and may possibly lead to new quantum information applications.

In this paper, we investigate the mixed states that, with respect to a given amount of purity, possess the maximal quantum violation of the Clauser-Horne-Shimony-Holt (CHSH) inequality for two-qubit systems, and of SI for three-qubit systems. Such states
are shown to be the optimal quantum resource in the context of quantum
nonlocal games: the CHSH game and Svetlichny's game, where the optimal states, compared to any classical strategy, are essential to
increase the probability of winning the games.

The  paper is organized as follows. In Sec. II, we describe the general definition of Svetlichny game from a family of Bell-type inequalities. We present and discuss the optimal state for the two-qubit CHSH game in Sec. III. Furthermore, we show the optimal state for the three-qubit Svetlichny's game in Sec. IV. We end the paper with a conclusion in Sec. V.

\section{The definition of multipartite Svetlichny game}

Nonlocal games serve as an equivalent way of describing tests for Bell-type inequalities\cite{Mermin,Mermin1,Peres,Brunner}. 
We first introduce some formal notations to define
Svetlichny's game in an $N$-party framework. We also suppose that one referee chooses an
$N$-bit question
\begin{eqnarray}
\mathcal{J}=i_1i_2\cdots i_N
\end{eqnarray}
uniformly from the
complete $N$-bit set, where $i_n=1,2 (n=1,...,N)$. He then sends $\mathcal{J}_1=i_1\dots i_j$ to
one group with $j$ players and $\mathcal{J}_2=i_{j+1}\dots i_N$ to
another with $N-j$ players. Each player $k\in\{1,2,...,N\}$ must
reply with a single bit $a_k$ as an answer
to the question $i_k$.
They win the game if and only if the answers
\begin{eqnarray}
\mathcal{A}=a_1\cdots a_N
\end{eqnarray}
satisfy the following criterion
\begin{eqnarray}
{\rm Mod}[\lfloor\frac{T}{2}\rfloor,2]=\bigoplus\limits_{k=1}^N
a_k,
\end{eqnarray}
with $T=T_1+T_2$ where $T_1$ and $T_2$ denote the times of bit ``1"
appeared in $\mathcal{J}_1$ and $\mathcal{J}_2$ respectively, and $\lfloor  x\rfloor$ refers to  the integer part of $x$.

With these notations, the winning probability of $N$ players is described as
\begin{eqnarray}
{\rm Pr}_N (\mbox{\rm win})=\frac{1}{2^N}\sum_\mathcal{J} P\biggr({\rm
Mod}[\lfloor\frac{T}{2}\rfloor,2]=\bigoplus\limits_{k=1}^Na_k\biggr).\label{PrN1}
\end{eqnarray}
Assuming that all players do not communicate with each another and that
the answer that each player returns is independent of any other players.
For a classical strategy, this implies that the joint probability is
separable, and we write
$P(a_k|i_k)=\frac{1}{2}(1+(-1)^{a_k} A_{k,i_k})$,
where $A_{k,i_k}\equiv\vec{\sigma}\cdot\vec{n}_{k,i_k}$ (with
$\vec{n}_{k,i_k}=\{\theta_{k,i_k},\phi_{k,i_k}\}$) is the observable of
the $k$-th qubit. The
winning probability (\ref{PrN1}) becomes
\begin{eqnarray}
{\rm Pr}_N (\mbox{\rm win})&=&\frac{1}{2^N}\sum_{\mathcal{J},\mathcal{A}}\delta_{\mathcal{J}\mathcal{A}}
P(a_1\cdots a_N|i_1\cdots i_N)\nonumber\\&=&\frac{1}{2^N}\sum_\mathcal{J}
\frac{1}{2}\biggr(1+(-1)^{\lfloor\frac{T}{2}\rfloor}A_{1,i_1}\cdots
A_{N,i_N}\biggr)\nonumber\\
&\equiv&\frac{2+\mathcal {S}_N}{4},\label{PrN2}
\end{eqnarray}
where $\mathcal {S}_N\leq1$ is just the form of the N-qubit SI defined in \cite{Svetlichny,us11}. Hence, $\delta_{\mathcal{J}\mathcal{A}}=1$ only when the answer $\mathcal{A}$ satisfies the  game criterion
for each question $\mathcal{J}$, otherwise $\delta_{\mathcal{J}\mathcal{A}}=0$.

In Eq.~(\ref{PrN2}), the equivalence between the quantum game and the
$N$-qubit SI is straightforward. Note that for given questions
$\mathcal{J}$ and answers $\mathcal{A}=a_1\cdots a_N$, the joint
probability $P(a_1\cdots a_N|i_1\cdots i_N)$ has non-zero
contributions only from the identity and the full correlation
$A_{1,i_1}\cdots A_{N,i_N}$. The other correlations do not contribute due to the symmetry of
the game criterion under permutation of any pair of players.

In fact, the $N$-qubit SI $\mathcal {S}_N\leq 1$ is a sum of
$2^{N-2}$ CHSH-type inequalities $\mathcal {I}_\alpha\leq 1$ (see
Ref.~\cite{us11} for details). This can be understood as follows. A
group of $j$ observers is denoted as a single party Alice and similarly the other group of
$N-j$ observers is denoted as Bob. The measuring results in
Alice's group are independent of those in Bob's, though observers in
 each group may be nonlocally
correlated.

As an example, the
winning probability (\ref{PrN2}) for $N=2$ can be expressed as
\begin{eqnarray}
{\rm Pr}_2&=&\frac{1}{2}+\frac{1}{8}(A_{11}A_{21}+A_{11}A_{22}+A_{12}A_{21}-A_{12}A_{22})\nonumber\\
&=&\frac{1}{4}(2+\mathcal {S}_2),
\end{eqnarray}
where $\mathcal {S}_2\leq1$ is the CHSH inequality. That is, the two-qubit Svetlichny's game is just the CHSH game. Since local realism requires $\mathcal {S}_2\leq1$, there is no classical strategy to win the CHSH game with probability exceeding $75\%$.
However, quantum mechanics can beat this bound. Consider the maximally entangled state $(\mid00\rangle+\mid11\rangle)/\sqrt{2}$ shared by Alice and Bob. There exists a quantum
strategy \cite{chsh69} such that the winning probability reaches
 $\frac{2+\sqrt{2}}{4}$, a quantum upper bound now known as Tsirelson's bound for the two-qubit system \cite{Tsirelson}.

Likewise, let us consider the three-qubit Svetlichny's game:
\begin{eqnarray}\label{pr3}
{\rm
Pr_3}&=&\frac{1}{4}(2+\mathcal {S}_3),
\end{eqnarray}
where \begin{eqnarray}\label{s3}
\mathcal {S}_3&=&\frac{1}{4}(A_{11}A_{21}A_{31}+A_{11}A_{21}A_{32}+A_{11}A_{22}A_{31}\nonumber\\
&&\;\;\;+A_{12}A_{21}A_{31}
-A_{11}A_{22}A_{32}-A_{12}A_{21}A_{32}\nonumber\\
&&\;\;\;-A_{12}A_{22}A_{31}-A_{12}A_{22}A_{32}),
\end{eqnarray}
with $\mathcal {S}_3\leq1$ being the three-qubit SI.
The hybrid local-nonlocal hidden variable models constrain the winning probability to be no more than $3/4$.
However, the genuine multipartite entanglement and quantum nonlocality
can be used to increase the probability. In fact, considering the GHZ state $\frac{1}{\sqrt{2}}(|000\rangle+|111\rangle)$ and choosing the proper measurement settings, the maximal amount of $\mathcal {S}_3$ in quantum system equals $\sqrt{2}$, which leads to that the probability of winning
the Svetlichny's game can attain $\frac{2+\sqrt{2}}{4}$.

In the following sections, we shall try to explore quantum games with a generic family of optimal states, in particular, the genuine maximally nonlocal mixed states (MNMS) characterized by maximizing the winning probability in Svetlichny's game for a given measure of mixedness of states. We shall use the linear entropy to quantify the mixedness, then study the maximal winning probability with knowledge of the CHSH and Svetlichny's inequalities.

\section{The  MNMS for the two-qubit CHSH game}

The normalized linear entropy is defined as~\cite{bose}
\begin{eqnarray}
\mathcal{E}_L(\rho)= \frac{d}{d-1}(1-\tr\rho^2),
\end{eqnarray}
with $d=2^N$. As an example, let us first consider the MEMS~\cite{MEMS} given by
\begin{eqnarray}
\rho^{\rm MEMS}=\left(  \begin{array}{cccc}
 g & & & \frac{\gamma}{2}\\
 &1-2g & &\\
    & & 0 &\\
    \frac{\gamma}{2} & & & g
\end{array}\right),
\end{eqnarray}
with $g=1/3$ for $\gamma\in[0,2/3)$, and $g=\gamma/2$ for $\gamma\in[2/3,1]$. Note that $\gamma$ quantifies the concurrence of the state.

\begin{figure}
\begin{center}
\includegraphics[width=0.4\textwidth]{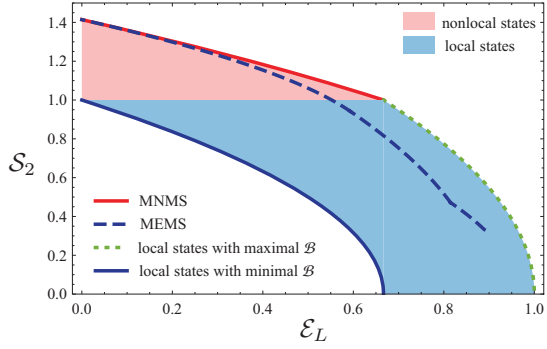} \\\vspace{-10pt}
 \caption{CHSH nonlocality versus linear entropy for two-qubit
 states.
 The red solid curve
$\mathcal {S}_2(\rho_2^{\rm MNMS})=\sqrt{2-3 \mathcal {E}_L /2}$ is for
the MNMS, which maximizes $\mathcal {S}_2(\rho_2^{\rm MNMS})$ for each
 value of $\mathcal {E}_L$ ($\mathcal {E}_L \leq 2/3$ so that
the CHSH inequality is violated). Pink and blue areas indicate,
respectively, arbitrary two-qubit nonlocal and local states in the
$\mathcal {E}_L$-$\mathcal {S}_2$ plane. The green dotted curve
$\mathcal {S}_2=\sqrt{3-3 \mathcal {E}_L}$ (corresponding to the state
(\ref{state2}) with $\vec{a}=\vec{b}=0$ and $c_3=0$) denotes the
maximal $\mathcal {S}_2$ for each value of $\mathcal {E}_L
> 2/3$. The blue solid curve $\mathcal {S}_2=\sqrt{1-3 \mathcal {E}_L
/2}$ (corresponding to the states $\rho=p |00\rangle \langle 00| +
(1-p) |11\rangle \langle 11|$) denotes the minimal $\mathcal {S}_2$ for
each fixed value of $\mathcal {E}_L \leq 2/3$. For comparison, we
also plot the MEMS denoted by the blue dashed
curve $\mathcal {S}_2(\rho_2^{\rm MEMS})=(\sqrt{2}+\sqrt{2-3 \mathcal
{E}_L})/2$ for $\mathcal {E}_L \in[0,16/27]$ and
$\mathcal {S}_2(\rho_2^{\rm MEMS})= \sqrt{25-27 \mathcal {E}_L -\min
\{1,3 (8-9 \mathcal {E}_L )/2 \} }/3$ for $\mathcal {E}_L
\in(16/27,8/9]$.}\label{fig1}\vspace{-20pt}
\end{center}
\end{figure}

Because we are interested in the region where $\langle\mathcal {S}_2\rangle$ can be violated by the state, we focus on the domain $\gamma\in[2/3,1]$, in which its linear entropy equals $\displaystyle \mathcal{E}_L=\frac{8}{3}\gamma(1-\gamma)$. By choosing the measurement directions in the $xy$-plane and $\phi_{11}=0, \phi_{12}=\pi/2, \phi_{21}=7\pi/4, \phi_{22}=\pi/4$ (see also the analysis below Eq. (\ref{Value3})), we see that its quantum maximum is  $\langle\mathcal {S}_2\rangle=2\sqrt{2}\gamma$. In other words, a relation between $\langle\mathcal {S}_2\rangle$ and $\mathcal{E}_L$ can be found as $\mathcal {S}_2(\rho_2^{\rm MEMS})=(\sqrt{2}+\sqrt{2-3 \mathcal{E}_L})/2$ for $\mathcal {E}_L \in[0,16/27]$ (see the blue dashed
curve in Fig. \ref{fig1}).

We next consider the Maximally Nonlocal Mixed States (MNMS).  In analogy to the MEMS which possesses the largest entanglement degree for a given linear entropy, we define the MNMS as a quantum mixed state that possesses the largest quantum violation of the CHSH inequality for a given linear entropy, and \emph{vice versa},  it is also a quantum state that possesses the largest linear entropy for a given quantum violation of the CHSH inequality. In the production of maximally entangled state, this state describes the output ports from a nonlinear crystal.  One way to obtain the generic form of MNMS involves the optimization of the violation of Bell-type inequalities. but this method may be very rather involved.
Instead, we consider a simpler yet rigorous method to
find the two-qubit MNMS.

In general, a two-qubit state can be written as
\begin{eqnarray}\label{state2}
\rho_2= \frac{1}{4} \biggr(I\otimes I + \sum_{i}r_{i}\sigma_{i} \otimes I  +
\sum_{j} s_{j}I\otimes \sigma_j \;\;\;\;\;\;\;\;\;\;\nonumber\\
+\sum_{m,n}^3 t_{mn} \sigma_m
\otimes \sigma_n \biggr),
\end{eqnarray}
where $\sigma_{i}$ is the Pauli matrix. The coefficients $(t_{mn})$ constitute a matrix $T$. The matrix $U=T^{T}T$ is symmetric, so  it can be diagonalized, with  $\lambda^{2}_{1},\lambda^{2}_{2},\lambda^{2}_{3}$. Here without loss of generality we have $|\lambda_{1}|\geq|\lambda_{2}|\geq|\lambda_{3}|$.

For state  (\ref{state2}), the linear entropy equals
\begin{eqnarray}\label{entropy}
 \mathcal {E}_L (\rho_2)&=& \frac{4}{3} (1-\tr\rho^2)\nonumber\\
 &=&1- \frac{1}{3}\biggr(\sum^3_{i=1}(r_{i}^2+s_{i}^2)+ \sum^3_{m,n=1}t_{mn}^2\biggr).
\end{eqnarray}

Our aim here is to find the MNMS that maximizes the violation of the CHSH inequality for a certain (\ref{entropy}).
As shown in Ref.~\cite{Horodecki1}, the maximal violation of the CHSH inequality with state (\ref{state2}) equals
\begin{equation}\label{MValue}
\mathrm{Max}\langle\mathcal {S}_2\rangle=\sqrt{\lambda_1^2+\lambda_2^2}.
\end{equation}
In order to maximize the linear entropy, according to Eq.(\ref{entropy}), one must minimize $\mathrm{Tr}\rho^2$. Hence it is reasonable to choose as many of the irrelevant coefficients as possible in (\ref{state2}) to be zero.
It is then shown that the violation determined by $\lambda_1$ and $\lambda_2$ are  related to $T$. For simplicity, suppose $U$ is diagonal. Hence, the simplest $U$ and $T$, except the zero matrix, read
\begin{eqnarray}
U=\left(  \begin{array}{ccc}
 \lambda_1^2 &0 &0\\
0 &\lambda_2^2 &0\\
   0 &0&0
\end{array}\right),\;\;\;\;T=\left(  \begin{array}{ccc}
 \lambda_1 &0 &0\\
0 &\lambda_2 &0\\
   0 &0&0
\end{array}\right).
\end{eqnarray}
Therefore, we obtain a matrix
\begin{eqnarray}\label{matrix}
M=\frac{1}{4}[\lambda_1\sigma_1\otimes\sigma_1+\lambda_2\sigma_2\otimes\sigma_2],
\end{eqnarray}
which maximally violates the CHSH inequality.

However, the matrix $M$ is not a physically allowed density matrix: the diagonal entries are zero with all nonzero elements of this matrix in the off-diagonal entries.
Given that $\tr\rho^2$ must be minimized,  in order to make $M$ physical, we can simply  add just four nonzero coefficients to the diagonal entries of $M$. By denoting these coefficients as $f_1$, $f_2$, $f_3$ and $f_4$, the new matrix $M'$ can be written as
\begin{equation}
M'=\left(  \begin{array}{cccc}
f_1&0 &0 &\frac{\lambda_1-\lambda_2}{4}\\
0& f_2&\frac{\lambda_1+\lambda_2}{4} &0 \\
0&\frac{\lambda_1+\lambda_2}{4} &f_3& 0 \\
\frac{\lambda_1-\lambda_2}{4}&0&0&f_4
\end{array}\right).
\end{equation}
Since matrix $M'$ is physical, it must satisfy the unit trace and positive definite requirements:
\begin{eqnarray}\label{positive}
\left\{
\begin{array}{l}
f_1+f_2+f_3+f_4=1, \\
f_1f_4\geq\frac{(\lambda_1-\lambda_2)^2}{16},  \\
f_2f_3\geq\frac{(\lambda_1+\lambda_2)^2}{16},
\end{array}
\right.
\end{eqnarray}
together with the condition that the  linear entropy must be maximized:
\begin{eqnarray}\label{entropy2}
\mathcal {E}_L (M')&=&\frac{4}{3}-\frac{1}{3}\biggr[f_1^2+f_2^2+f_3^2+f_4^2+\frac{(\lambda_1-\lambda_2)^2}{16}\nonumber\\
&&\;\;\;\;\;\;\;\;\;\;\;\;+\frac{(\lambda_1+\lambda_2)^2}{16}\biggr]\nonumber\\
&\leq&\frac{4}{3}-\frac{1}{3}\biggr[2f_1f_4+2f_2f_3+\frac{(\lambda_1-\lambda_2)^2}{16}\nonumber\\
&&\;\;\;\;\;\;\;\;\;\;\;\;+\frac{(\lambda_1+\lambda_2)^2}{16}\biggr].
\end{eqnarray}
The sign of equality can be only achieved when $f_1=f_4$ and $f_2=f_3$, at which Eqs.~(\ref{positive}) lead to $\lambda_1=1$.
Finally, we obtain the two-qubit MNMS
\begin{eqnarray}\label{MNMS2}
\rho_2^{\rm MNMS}=\left(  \begin{array}{cccc}
\frac{1-\lambda_2}{4}&0 &0 &\frac{1-\lambda_2}{4}\\
0& \frac{1+\lambda_2}{4}&\frac{1+\lambda_2}{4} &0 \\
0&\frac{1+\lambda_2}{4} &\frac{1+\lambda_2}{4}& 0 \\
\frac{1-\lambda_2}{4}&0&0&\frac{1-\lambda_2}{4}
\end{array}\right).
\end{eqnarray}
This ends the proof.

The form of (\ref{MNMS2})
may seem somewhat abstract. However, it is very interesting to note
that this MNMS can be rewritten into a new form with
 intuitive physical meaning, i.e.,
\begin{eqnarray}
\rho^{\rm MNMS}_2=\frac{1+\gamma}{2}\rho_1+\frac{1-\gamma}{2}\rho_2,\label{MNMSS}
\end{eqnarray}
where $\gamma=\lambda_2\in[-1,1]$ represents a mixture of two orthogonal
states $\rho_i=|\psi_i\rangle\langle\psi_i|$, with $|\psi_1\rangle =
\frac{1}{\sqrt{2}}(|00\rangle+|11\rangle)$ and
$|\psi_2\rangle =
\frac{1}{\sqrt{2}}(|01\rangle+|10\rangle)$.
So the two-qubit MNMS  can be
considered as a imperfect Bell state  with random spin flipping, where $\gamma$ represents a parameter to describe such a  flip. In particular, when $\gamma=1,$ Eq.~(\ref{MNMSS}) equals the maximally entangled state; when $\gamma=0,$ it becomes a separable state.

For this state, note that the maximum of $\langle\mathcal {S}_2\rangle$ equals $\sqrt{1+\gamma^2}$ by choosing proper directions, while  $\max\mathcal
{E}_L=1-\frac{1}{3}\bigr(1+2\gamma^2\bigr)$. When the MNMS is applied to the CHSH game, the winning probability reaches ${\rm Pr}^{\rm MNMS}_2(\mbox{\rm win})=
(2+\sqrt{1+\gamma^2})/4$, which ranges from $3/4$ to $(2+\sqrt{2})/4(\approx0.8535)$, the latter being the quantum upper bound in the game.

\begin{figure}
\begin{center}
\includegraphics[width=0.4\textwidth]{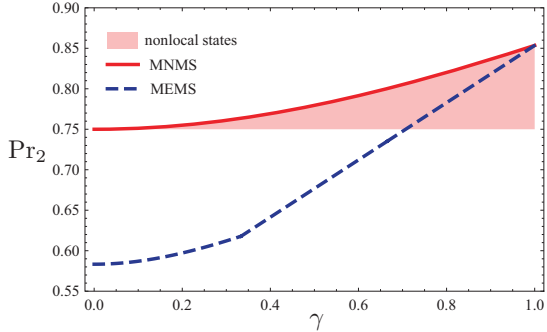} \\\vspace{-10pt}
 \caption{The quantum winning probability in the CHSH game versus concurrence
 $\gamma$ for the MNMS (red
solid curve), MEMS (blue dashed curve), and arbitrary states (pink region) that violate the CHSH inequality. Here the concurrence depicts degree of entanglement of two qubits.
If the quantum state
shared by Alice and Bob is the MNMS, the quantum winning probablity
equals ${\rm Pr^{MNMS}_2}= (2+\sqrt{1+\gamma^2})/4$, while for the MEMS the
winning probability is found to be ${\rm Pr^{MEMS}_2}= (6+\sqrt{1+18
\gamma^2-\min\{1,9 \gamma^2\}})/12$ for $\gamma\in[0,2/3]$ and ${\rm
Pr^{MEMS}_2}= (2+\sqrt{2}\gamma)/4$ for $\gamma\in(2/3,1]$. It is
clearly shown that ${\rm Pr^{MNMS}_2}$ surpasses ${\rm Pr^{MEMS}_2}$
except $\gamma=1$, at which the  MNMS and MEMS are the Bell states
resulting in the largest probability $\frac{1}{4}(2+\sqrt{2})$.
}
\label{fig2}\vspace{-25pt}
\end{center}
\end{figure}

In Figure~\ref{fig1}, we consider arbitrary two-qubit
states and plot  the $\mathcal {S}_2-\mathcal
{E}_L$ plane. The MNMS (see the red solid curve) serves as the optimal state that maximizes
$\mathcal {S}_2$ for a fixed value of $\mathcal
{E}_L$. We also plot the MEMS~\cite{MEMS,MEMS1} (see the blue dashed line) for comparison.
Apparently, the MNMS does not overlap with the MEMS and is thus a distinct family of states.
In Fig.~\ref{fig2}, we plot the winning probability with MNMS and MEMS in the CHSH game. It is clearly shown that the MNMS serves as an upper bound of quantum strategy.

\section{The genuine MNMS for three-qubit Svetlichny's game}

The genuine MNMS is defined as the optimal state in Svetlichny's game for a given value of linear entropy. According to Eq. (\ref{pr3}), such a  state must maximally violate three-qubit SI for a given value of linear entropy. To this end, we note that similar to Eq. (\ref{state2}), any three-qubit state, up to local unitary operations, can be expressed in terms of Pauli matrices, originally defined in Ref. \cite{Horodecki1,Horodecki2}. However, the form of the three-qubit density matrix consists of $63$ coefficients, a far more complicated situation  than that of a two-qubit state (\ref{state2}).

For a three-qubit state $\rho_3$, the average value of SI reads
\begin{equation}\label{Value3}
\langle\mathcal {S}_3\rangle=\mathrm{Tr}(\rho_3\mathcal {S}_3).
\end{equation}
The computation of the Tsirelson bounds is in general not an obvious task. However, the task could be somewhat simplified for some particular cases: for instance when the SI can be expressed as a sum of CHSH-type inequalities~\cite{chen3}. In order to obtain the maximum value for the quantum system, numerical results show that it is enough to consider measurement settings $A_{ij}$ confined within the $xy$-plane, i.e., $A_{ij}\equiv\sigma_{ij}=\cos{\phi_{ij}}\sigma_x+\sin{\phi_{ij}}\sigma_y$. The Svetlichny operator $\mathcal{S}_3$ is then a matrix with nonzero terms only in off-diagonal entries, i.e., $\langle\mathcal{S}_3\rangle$ only depends on off-diagonal terms of $\rho_3$.

As discussed in the previous section for two qubits,  the problem reduces to finding a genuine MNMS that possesses the maximal linear entropy from the set of states with the same violation of SI.
To this end, we note that the normalized linear entropy of $\rho_3$ equals $\mathcal {E}_L (\rho_3)= \frac{8}{7} (1-\tr\rho_3^2)$. Maximizing this entropy is equivalent to
minimizing the quantity $\mathrm{Tr}(\rho_3^2)$.

Hence, for a fixed violation $\langle\mathcal{S}_3\rangle$, one chooses all irrelevant terms of the density matrix to be zero, while keeping the conditions of positive semi-definiteness and trace unity satisfied.
We then obtain a \emph{necessary} form of the MNMS (see also \cite{chen2}):
\begin{equation}\label{opt density11}
\rho=\left(  \begin{array}{cccccccc}
\rho_{11}&0 &0 &0 &0&0 &0&\rho_{18}\\
0&\rho_{22} &0 &0 &0&0 &\rho_{27}&0\\
0&0 &\rho_{33} &0 &0&\rho_{36}&0&0\\
0&0 &0 &\rho_{44} &\rho_{45}&0 &0&0\\
0&0 &0 &\rho_{54} &\rho_{55}&0 &0&0\\
0&0 &\rho_{63} &0 &0&\rho_{66} &0&0\\
0&\rho_{72} &0 &0 &0&0 &\rho_{77}&0\\
\rho_{81}&0 &0 &0 &0&0 &0&\rho_{88}
\end{array}\right),
\end{equation}
where
\begin{eqnarray}
\rho_{mm}=\rho_{nn}=\sqrt{\|\rho_{mn}\|^2},\;\;\;\;n=9-m.
\end{eqnarray}

The justification of the $X$-shaped form of Eq. (\ref{opt density11}) is as follows: The anti-diagonal entries can in general be nonvanishing, since with all measurement directions confined within the $xy$-plane, such entries determine Tsirelson's bound of SI. The remaining entries should then be chosen as zeros in order to minimize $\mathrm{Tr}(\rho^2)$; however, a matrix with only nonzero anti-diagonal entries is not a physical state. To get a physical state, diagonal entries with proper values are thus necessary to make the matrix both have a unit trace and being semi-positive definite. Thus, the MNMS can be shown to be restricted to an $X$-shaped form, as shown in Eq. (\ref{opt density11}).

Finding the genuine MNMS from Eq.(\ref{opt density11}) is then equivalent to solving an optimization problem of finding a matrix that minimizes $\mathrm{Tr}(\rho^2)$ (or maximizes $\mathcal {E}_L$) for each value of $\langle \mathcal{S}_3\rangle$:
\begin{equation}
\rho^{\rm MNMS}_3\left\{ \begin{aligned}
          &{\rm maximize}\;\mathcal {E}_L {\rm \; for \;each\;}\langle \mathcal{S}_3\rangle \\
          &s.t.:\; {\rm semi-positivity,\;unit \; trace}
 \end{aligned} \right.
\end{equation}
or that maximizes $\langle \mathcal{S}_3\rangle$ for each value of $\mathcal {E}_L$:
\begin{equation}
\rho^{\rm MNMS}_3\left\{ \begin{aligned}
          &{\rm maximize}\;\langle \mathcal{S}_3\rangle {\rm \; for \;each\;}\mathcal {E}_L \\
          &s.t.:\;{\rm semi-positivity,\;unit \; trace}
 \end{aligned} \right.
\end{equation}

We list the results as follows:
\begin{eqnarray}
&\rho_{11}=f_1\in[\frac{1}{8},\frac{1}{2}],& \nonumber\\
&\rho_{22}=\rho_{33}=\rho_{44}=f \in[0, \frac{1}{8}],&\\
&f_1+3f=\frac{1}{2},&\nonumber
\end{eqnarray}
which leads to the desired three-qubit genuine MNMS
\begin{equation}\label{opt density}
\rho_{3}^{\rm MNMS}=\left(  \begin{array}{cccccccc}
f_1&0 &0 &0 &0&0 &0&f_1\\
0&f &0 &0 &0&0 &f&0\\
0&0 &f &0 &0&f &0&0\\
0&0 &0 &f &f&0 &0&0\\
0&0 &0 &f &f&0 &0&0\\
0&0 &f &0 &0&f &0&0\\
0&f &0 &0 &0&0 &f&0\\
f_1&0 &0 &0 &0&0 &0&f_1
\end{array}\right).
\end{equation}

Let us now take a closer look at the results on $\rho_{3}^{\rm MNMS}$. For this state, the following settings
\begin{eqnarray}
&\phi_{11}=\phi_{21}=-\theta,&\nonumber\\
& \phi_{12}=\phi_{22}=\phi_{31}=\theta,&\nonumber\\
 & \phi_{32}=\pi-\theta,&\\
  & \theta=\arccos{\sqrt{\frac{1-8f}{2-24f}}},&\nonumber
\end{eqnarray}
can be used to achieve the quantum maximum value for SI:
\begin{eqnarray}\label{maximal violation SI}
\langle\mathcal {S}_{3}^{\mathrm{Max}}\rangle&=&\frac{(1-8f)^{\frac{3}{2}}}{{(\frac{1}{2}-6f)^{\frac{1}{2}}}}\qquad\;\;{\rm for}\;\; 0\leq f\leq \frac{1}{16},\\
\langle\mathcal {S}_{3}^{\mathrm{Max}}\rangle&=&1 \qquad\qquad\qquad\;\;{\rm for}\;\;\frac{1}{16}\leq f\leq \frac{1}{8}.
\end{eqnarray}
Note that the genuine MNMS equals the standard GHZ state at $f=0$, achieving the maximal violation $\sqrt{2}$.

\begin{figure}
\begin{center}
\includegraphics[width=0.4\textwidth]{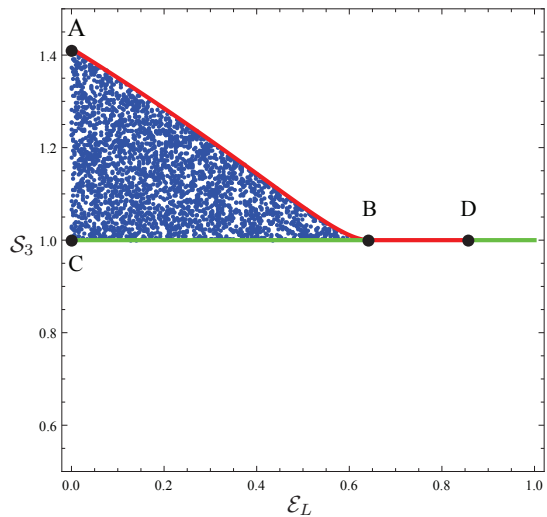} \\
 \caption{The genuine multipartite nonlocality versus the linear entropy for three-qubit states. Point A corresponds to the GHZ state $(|000\rangle+|111\rangle)/\sqrt{2}$, B to the MNMS (\ref{opt density}) at $\mathcal {E}_L=9/14$, C to $\cos\pi/8|000\rangle+\sin\pi/8|111\rangle$, and D to the MNMS (\ref{opt density}) at $\mathcal {E}_L=6/7$. The red curves AB and BD  correspond to the maximal violation of SI with state (\ref{opt density}) for each $\mathcal {E}_L$ (see also Eqs.~(\ref{maximal violation SI1}) and (\ref{maximal violation SI11})). The solid line crossing points $B$, $C$ and $D$ is the classical bound. The blue points represent a great number of randomly chosen states.}
\label{fig3}
\end{center}
\end{figure}

The linear entropy of the genuine MNMS equals
\begin{eqnarray}
\mathcal {E}_L (\rho_3^{\rm MNMS})=\frac{96}{7}f(1-4f).
\end{eqnarray}
The quantum maximum value of SI can then be rewritten with the linear entropy in its argument:
\begin{eqnarray}\label{maximal violation SI1}
\langle\mathcal {S}_{3}^{\mathrm{Max}}\rangle&=&\frac{(1-\frac{1}{6}w)^{\frac{3}{2}}}{({\frac{1}{2}-\frac{1}{8}w})^{\frac{1}{2}}}\;\;\;\;\;
 {\rm for} \;\; 0\leq \mathcal {E}_L< \frac{9}{14},\\
\langle\mathcal {S}_{3}^{\mathrm{Max}}\rangle&=& 1 \qquad\qquad\qquad  {\rm for} \;\;\frac{9}{14}\leq \mathcal {E}_L\leq \frac{6}{7}.\label{maximal violation SI11}
\end{eqnarray}
with $w=6-\sqrt{6}\sqrt{6-7\mathcal {E}_L}$ (see Fig.~\ref{fig3} for a graphic illustration).

Obviously, when MNMS is applied to three-qubit Svetlichny's game, the winning probability ranges from $\displaystyle \frac{3}{4}$ to $\displaystyle \frac{(2+\sqrt{2})}{4}$, similar to the two-qubit MNMS case. A major difference here is that there is a line of maxima for SI as the entropy increases between ${9}/{14}$ and ${6}/{7}$ (see the green solid line $BD$ in Fig.~\ref{fig3}).
In Fig. \ref{fig3}, we plot $\langle\mathcal {S}_{3}^{\mathrm{Max}}\rangle$ versus $\mathcal {E}_L (\rho_3^{\rm MNMS})$ with a great number of randomly chosen states, to confirm our analytic results.

In fact, the genuine MNMS can be understood as a standard GHZ state subjected to classical errors, namely, $X$-type errors, with each spin undergoing bit flip with equal probability. To be specific, if we first prepare a pure GHZ state for a tripartite spin-1/2 system
\begin{eqnarray}
\mid \psi_1\rangle=\frac{1}{\sqrt{2}}(\mid000\rangle+\mid111\rangle)
\end{eqnarray}
and allow it to undergo a noisy channel such that the computational basis states $|0\rangle$ and $|1\rangle$ are flipped with an equal probability, say $f$, one spin at a time (or equivalently, two spins at a time) so that  the initial state becomes one of the three flipped states
\begin{eqnarray}
\mid \psi_2\rangle=\frac{1}{\sqrt{2}}(\mid100\rangle+\mid011\rangle),\\
 \mid \psi_3\rangle=\frac{1}{\sqrt{2}}(\mid010\rangle+\mid101\rangle),\\
   \mid \psi_4\rangle=\frac{1}{\sqrt{2}}(\mid001\rangle+\mid110\rangle).
\end{eqnarray}
with the same probability.
The resultant state through the channel can then be described by
\begin{eqnarray}\label{opt density2}
\rho_{3}^{\mathrm{MNMS}}=2f_1\rho_1+f\rho_2+f\rho_3+f\rho_4,
\end{eqnarray}
where $\rho_i=|\psi_i\rangle\langle\psi_i|$, yielding nothing but the MNMS (\ref{opt density}). It is apparent that such an analysis also applies to the two-qubit case (\ref{MNMSS}).

The MNMS serves as a good approximation of maximally entangled states that undergo classical errors in experimental preparation of states. We would like to stress that for MNMS,  the influence of environment on pure quantum states is different from that for the MEMS, for which nonclassical correlations are quantified in terms of entanglement of formation needed for creating the given state~\cite{eof}. We believe that the proposed notion of MNMS may therefore be more practical than MEMS in many quantum information processes where classical errors are dominant.


\section{conclusion}
In summary, we have derived the optimal states, the MNMS, which provide maximal violation of the CHSH and three-qubit Svetlichny's inequality for a given mixedness of states.
It has been clearly shown that the MNMS is distinct from the MEMS, in that they give comparatively different curves in the $\mathcal {S}-\mathcal
{E}_L$ plane.

For the two-qubit system, the upper bound of linear entropy of the MNMS can reach
${2}/{3}$, within which the quantum strategy will have a chance to beat its classical counterpart, while for the three-qubit system the value equals ${9}/{14}$ and, as a qualitative difference from the two-qubit case, there exists a terrace in the $\mathcal {S}-\mathcal
{E}_L$ plane for $\mathcal{E}_L\in[9/14,6/7]$ (see Figs.~\ref{fig1} and \ref{fig3} for comparison).

Moreover, we have also pointed out that the MNMS can be a good representation of maximally entangled states that have undergone  $X$-type errors, i.e.  local spin flips.
We also see that the two-qubit and three-qubit MNMS are tolerant against white noise, serving as  a valuable resource for quantum information and computation protocols involving Bell-type nonlocality, such as quantum nonlocal games, Bell's-theorem-based quantum cryptography, Bell's-theorem-based random number generator, etc.  We expect that our method  may
cast a new perspective for understanding quantum games for general mixed states scenarios.  Further questions, like the proofs of MNMS for arbitrarily multiple high-dimensional systems, remain an open question and we hope that we can investigate these issues at length  in the future.

\section*{ACKNOWLEDGMENTS}

J.L.C. is supported by the National Basic Research Program (973 Program) of China under Grant No. 2012CB921900 and the Natural Science Foundations of China
(Grant Nos. 11175089 and 11475089). C.R.thanks Chern Institute of Mathematics for invited visiting and acknowledges supported by Youth Innovation Promotion Association (CAS) No.2015317, Natural Science Foundations of Chongqing (No.cstc2013jcyjC00001, cstc2015jcyjA00021, Y31Z0P0W10) and The Project-sponsored by SRF for ROCS-SEM (No.Y51Z030W10). H.Y.S. acknowledges the support by Institute for Information and Communications Technology Promotion (IITP), Daejeon, Republic of Korea.
This work is also partly supported by the National Research Foundation and the Ministry of Education, Singapore.


\begin{thebibliography}{99}
\bibitem{EPR} A. Einstein, B. Podolsky, N. Rosen, Phys. Rev. \textbf{47}, 777 (1935).
\bibitem{Acin07} A. Ac\'{i}n, Phys. Rev. Lett. \textbf{98}, 230501 (2007).
\bibitem{Pironio10}S. Pironio \emph{et al.}, Nature \textbf{464}, 1021-1024 (2010).
\bibitem{Bell64} J. S. Bell, \emph{Physics} (Long Island City, N.Y.) \textbf{1}, 195 (1964).
\bibitem{chsh69}J. Clauser, M. Horne, A. Shimony, R. Holt, Phys. Rev. Lett. \textbf{23}, 880 (1969).
\bibitem{MABK1} N. D. Mermin, Phys. Rev. Lett. \textbf{65}, 1838 (1990).
\bibitem{MABK2}M. Ardehali, Phys. Rev. A \textbf{46}, 5375 (1992).
\bibitem{MABK3}A. V. Belinskii, D. N. Klyshko, \emph{ Phys. Usp.} \textbf{36}, 653-693 (1993).
\bibitem{WWZB1}M. \.{Z}ukowski, C. Brukner, Phys. Rev. Lett. \textbf{88}, 210401 (2002).
\bibitem{WWZB2}R. F. Werner, M. M. Wolf, Phys. Rev. A \textbf{64}, 032112 (2001).
\bibitem{Svetlichny}G. Svetlichny, Phys. Rev. D \textbf{35}, 3066 (1987).
\bibitem{Svetlichny2}M. Seevinck, G. Svetlichny, Phys. Rev. Lett. \textbf{89}, 060401 (2002).
\bibitem{Svetlichny3}D. Collins, N. Gisin, S. Popescu, D. Roberts, V. Scarani, Phys. Rev. Lett. \textbf{88}, 170405 (2002).
\bibitem{Gisin11} J. D. Bancal, N. Brunner, N. Gisin, Y. C.  Liang, Phys. Rev. Lett. \textbf{106}, 020405(2011).
\bibitem{us11} J. L. Chen, D. L. Deng, H . Y. Su, C. F.  Wu, C. H. Oh, Phys. Rev. A \textbf{83}, 022316 (2011).
\bibitem{acin2012} L. Aolita, R. Gallego, A. Cabello, and A. Ac\'in, Phys. Rev. Lett. \textbf{108}, 100401 (2012).
\bibitem{lee2013} S.-W. Lee, M. Paternostro, J. Lee, and H. Jeong, Phys. Rev. A \textbf{87}, 022123 (2013).
\bibitem{adesso2014} G. Adesso and S. Piano, Phys. Rev. Lett. \textbf{112}, 010401 (2014).
\bibitem{MEMS} W. J. Munro, D. F. V. James, A. G. White, P. G. Kwiat, Phys. Rev. A \textbf{64}, 030302(R) (2001).
\bibitem{MEMS1} T. C. Wei, K. Nemoto, P. M. Goldbart,P. G. Kwiat,W. J. Munro, F. Verstraete, Phys. Rev. A \textbf{67}, 022110 (2003).
\bibitem{MDMS} F. Galve, G. L. Giorgi, R. Zambrini, Phys. Rev. A \textbf{83}, 012102 (2011).
\bibitem{Mermin} N. D. Mermin, Am. J. Phys. 58, 731 (1990).
\bibitem{Mermin1} N. D. Mermin, Phys. Rev. Lett. 65, 3373 (1990).
\bibitem{Peres} A. Peres, Phys. Lett. A 151, 107 (1990).
\bibitem{Brunner} N. Brunner and N. Linden, Nat. Commun. 4, 2057 (2013).
\bibitem{Tsirelson} B. S. Tsirelson, Lett. Math. Phys. \textbf{4}, 93-100 (1980).
\bibitem{bose} S. Bose and V. Vedral, Phys. Rev. A \textbf{61}, 040101(R) (2000).
\bibitem {Horodecki1} R. Horodecki, P. Horodecki, M. Horodecki, Phys. Lett. A \textbf{200}, 340 (1995).
\bibitem {Horodecki2} R. Horodecki, P. Horodecki, M. Horodecki, K. Horodecki, Rev. Mod. Phys. \textbf{81}, 865 (2009).
\bibitem {chen3} J. L. Chen, D. L. Deng, H. Y. Su, C. Wu, and C. H. Oh, Phys. Rev. A \textbf{83}, 022316 (2011).
\bibitem {chen2} C. L. Ren, H. Y. Su, Z. P. Xu, C. F. Wu, J. L. Chen, Sci. Rep. \textbf{5}, 13080 (2015).
\bibitem {eof} C. H. Bennett, D. P. DiVincenzo, J. A. Smolin, and W. K. Wootters, Phys. Rev. A \textbf{54}, 3824 (1996).
\end{thebibliography}
\end{document}